# Clustering-based Anomaly Detection for microservices

*Roman Nikiforov*

*DINO Systems, Saint Petersburg, Russia*

*roman.nikiforov@nordigy.ru*

## *Abstract*

*Anomaly detection is an important step in the management and monitoring of data centers and cloud computing platforms. The ability to detect anomalous virtual machines before real failures occur results in reduced downtime while operations engineers urgently recover malfunctioning virtual machines, efficient root cause analysis, and improved customer optics in the event said malfunction lead to an outage. Virtual machines could fail at any time, whether in a lab or production system. If there is no anomaly detection system, and a virtual machine in a lab environment fails, the QA and DEV team will have to switch to another environment while the OPS team fixes the failure. The potential impact of failing to detect anomalous virtual machines can result in financial ramifications, both when developing new features and servicing existing ones. This paper presents a model that can efficiently detect anomalous virtual machines both in production and testing environments.*

## 1. Introduction

The task of processing large amounts of data to find deviations is not trivial. The common approach of dividing a product into microservices only makes it harder. More services equal more interactions within a system and leads to even more data being collected on a day-to-day basis. Each microservice is basically a separate Virtual Machine (VM) with its own metrics in regards to central processor unit (CPU) load, memory usage, disk usage, etc. Since a failure of even a single component in microservices-based architecture might cause a malfunction of all systems in general, it is very important to be able to catch such components in a pre-failure state. The problem is especially relevant to development and testing environments, which imitate a production model but without a full high-availability (HA) representation due to the need to save resources. A failure on any particular VM has far reaching effects as it causes a temporary increase in load on other VMs, in particular any micro-service.

This paper will present a way of detecting any outlying VMs using density-based spatial clustering of applications with noise (DBSCAN)[1]. Once the outliers are identified, statistical-based review will be performed to verify results.

In a production microservices based model, each service usually operates as a part of cluster, which might contain several units to tens of units. Each of these units is a VM running the same type of software which should behave similarly to all other VMs of the cluster. From the development side, having multiple testing environments imitating a production model leads to a set of VMs within a data center performing similar tasks but within different logical clusters. Trying to detect an anomalous component within a single testing environment which consists of different micro-services based only on data from the VMs from that particular environment is a poor approach since different components perform different tasks. The metrics being collected won't represent issues at a micro-services level and will be nothing more than noise.

Section 2 of this article gives a brief overview of existing methods for anomaly detection, such as Multivariate Adaptive Statistical Filtering (MASF)[2], Multivariate Gaussian (normal) distribution[3], and support vector machine (SVM) based anomaly detection algorithms[4].Section 3 of this article will introduce a way of picking the right metrics for analysis and describe the presented algorithm in detail. Section 4 covers the results of experiments that were conducted. Section 5 is a conclusion of this article and a glance into future work.

## 2. Background

In their work, J.P. Buzen and A.W. Shum from BGS Systems, Inc. have introduced a way of analyzing computer system performance called Multivariate Adaptive Statistical Filtering (MASF). This method is based on tailored reference sets containing previously measured values that were collected under typical operating conditions. There are several problems of using MASF in a system built from microservices. First, it is hard to mark an operational condition at any point in time as typical vs atypical since there is no information available on whether a VM is under load that is caused by a planned action, such as auto-tests run in a testing environment or a temporary surge in user activity on a production system, or if this load is unexpected. Second, there is no information on whether idle VMs didn't have any load due to no proper requests at the time, some failure in a system, or this machine simply didn't receive any requests from another component that should have sent requests. Third, the MASF model collects statistical data for a period of time and marks it 'normal'. This data is then used as a reference and compared to collected data in a future time period. If there wasn't a proper load during the reference data collection window it could not be used as a standard later on when a load might (would) be presented. Lastly, MASF generally relies on the assumption that the collected data has a Gaussian (normal) distribution which is not true in a number of cases.

Figure 1 displays some examples of CPU usage distribution among all hosts running the same component. It is easy to see that the distribution does not conform to Gaussian, let alone Memory usage statistics presented later on Figure 2. This fact makes a use of MASF as well as any other Gaussian-based model inexpedient.

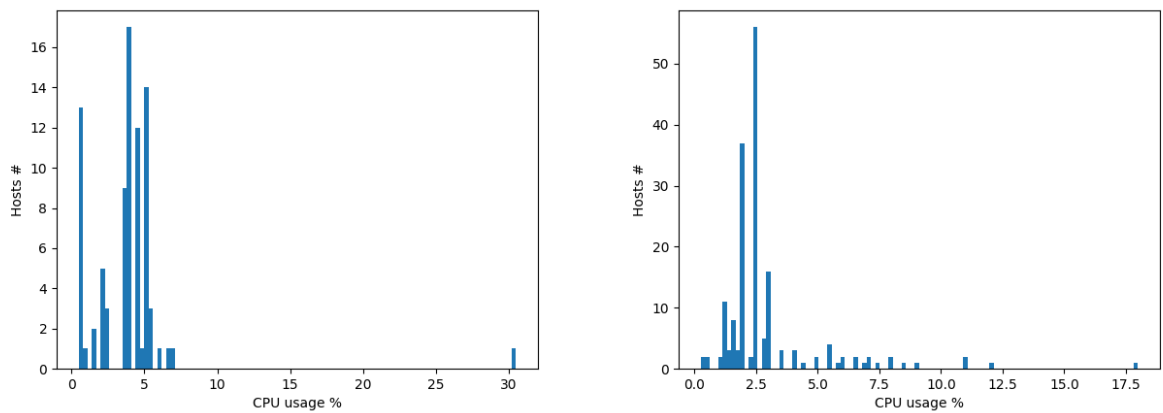

Figure 1. CPU usage distribution among hosts running same component in different LAB environments

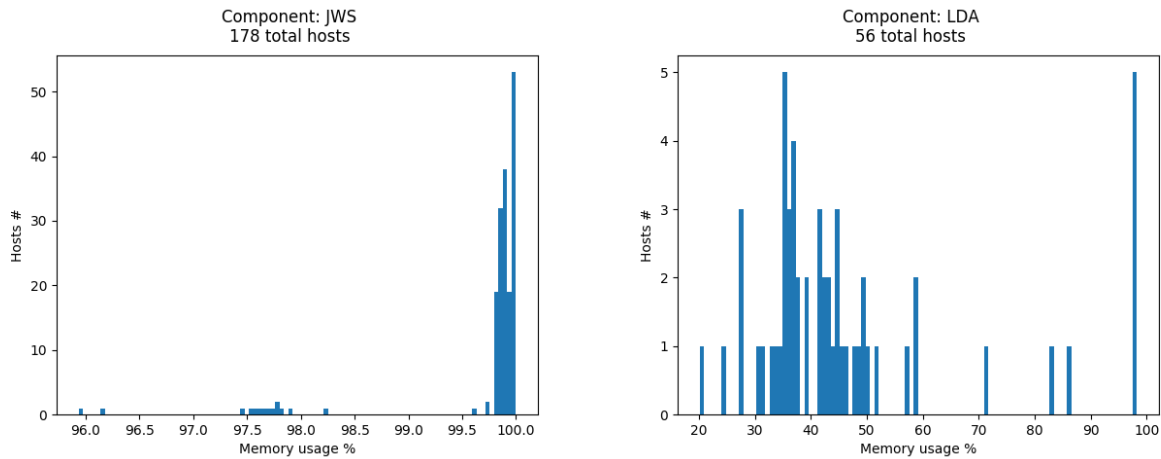

Figure 2. Memory usage distribution

GuiPing Wang and JiaWei Wang present An Anomaly Detection Framework for Detecting Anomalous Virtual Machines under Cloud Computing Environment (EeAD) based on several SVM-based anomaly detection algorithms, including C-SVM, OCSVM, multi-class SVM and imbalanced SVM. One of the main ideas in their work is a partitioning of all VMs in Cloud platform into several monitoring domains based on similarity in running environment. This helps to segregate virtual machines before further analysis. However, using supervised learning methods requires a knowledge of each VM's state at the moment when statistics are being collected. In order to do this, you either have to assume all VMs are not anomalous, which is a bad approach for multiple reasons, or manually check every single VM for an anomaly. The first approach is not good because in supervised learning binary classification you want to have both positive and negative examples. The second approach requires a lot of time to manually record the state of each VM. One possible way to use the first approach is to deliberately create anomalous examples by anomaly-injection techniques, but without preliminary checking of all other VMs one can never be sure that all of them are not anomalous.

By itself, the idea of using a method based on a supervised learning algorithm is good, but it requires very careful data analysis before making a training example from the collected metrics. The model presented in this article, if used for a long enough period of time, might be able to collect enough data to be used in supervised learning algorithm later on.

## 3. Detection Principles

The method for anomaly detection presented in this article is based on the fundamental principle of organizing all the VMs in the system into multiple domains with several nodes in each domain based on the component each node is running. Since the same components across the system should behave similarly, doing the same tasks and running the same software, you want to group them. E.g. components that are responsible for routing HTTP requests might have higher CPU usage and lower input/output operations per second (iops), whereas components running databases would have higher read operations per second and high memory usage. Some components are only routing the requests and some of them are making calculations during data processing, so it is a wise idea to combine the same components into a domain.

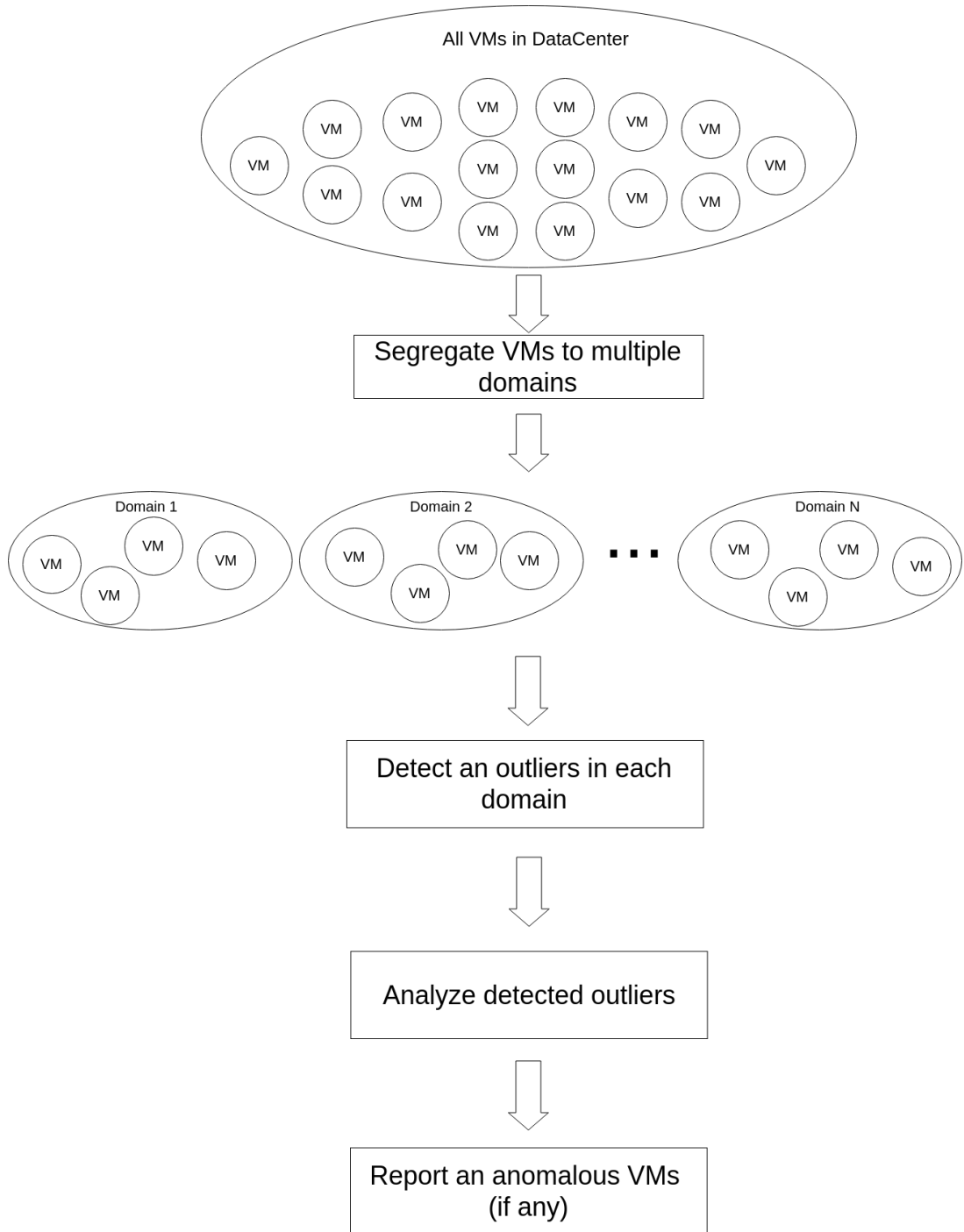
Figure 3. The detection principles

As shown on Figure 3, VMs are segregated into multiple domains based on the components they are running. This task does not require any significant amount of calculation since every VM is already organized according to the component that it is running, which requires only O(N) time complexity algorithm. A set of performance metrics for each VM is also collected at this time. In the next step each domain is being examined with DBSCAN algorithm in order to find any outliers present in the domain. DBSCAN has a complexity of O(N*log N) so running this in each domain will result in O(M * N * log N) where M is the number of domains and N is the number of VMs in each domain.

For each VM, the following metrics are collected: CPU utilization in percent, memory usage in percent, the number of input operations per second, and the number of output operations per second. This list may be extended by any other significant metrics. This set represents a X ∈ $R^n$ where $n$ is the number of metrics. The metrics set can be formalized as:

$$X = \begin{bmatrix} x_1 \\ x_2 \\ \vdots \\ x_n \end{bmatrix}$$

Where $x_i$ is a real number representing the particular metric value for a particular VM.

A domain performance metrics set is a set of sets where each column is a vector of a particular VM's metrics values. It can be formalized by the following matrix $Y$:

$$Y = \begin{bmatrix} X_1 X_2 \cdots X_m \end{bmatrix}, Y_{n \times m} = \begin{bmatrix} x_{1,1} & x_{1,2} & x_{1,3} & \cdots & x_{1,m} \\ x_{2,1} & x_{2,2} & x_{2,3} & \cdots & x_{2,m} \\ \vdots & \vdots & \vdots & \ddots & \vdots \\ x_{n,1} & x_{n,2} & x_{n,3} & \cdots & x_{n,m} \end{bmatrix}$$

Where $n$ is a number of metrics per VM and $m$ is a number of VMs in a domain.

DBSCAN clustering operates with a following terms:

- A point $p$ is a core point if at least *minPts* points are within distance $\varepsilon$ ($\varepsilon$ is the maximum radius of the neighborhood from $p$) of it (including $p$). Those points are said to be *directly reachable* from $p$.

- A point $q$ is directly reachable from $p$ if point $q$ is within distance $\varepsilon$ from point $p$ and $p$ must be a core point.

- A point $q$ is reachable from $p$ if there is a path $p_1, ..., p_n$ with $p_1 = p$ and $p_n = q$, where each $p_{i+1}$ is directly reachable from $p_i$ (all the points on the path must be core points, with the possible exception of $q$).

- All points not reachable from any other point are outliers.[5]

Based on the above two parameters needs to be set: $\varepsilon$ and *minPts*.

For the purpose of the anomaly detection method presented in this article, *minPts* has been chosen as a number of samples in a domain divided by 3. This means that at least one third of the total samples out of one domain should be able to establish a cluster. This number was picked based on the idea of having 3 possible clusters maximum, so in perfect conditions one of them might gather VMs being idle at the moment, the second one would have VMs under an extensive load, and the third one might consist of machines with an average load. Although in practice it is expected that all VMs would create a single cluster with or without individual outliers.

$$minPts = \frac{m}{3}$$

Second parameter ε, which is a maximum distance from point *p* to any other point to be considered as a cluster. In order to define this parameter, scaling needs to be applied to the *Y* matrix to shrink all metrics into a range of [0;1]. Now a maximum value within every set of the same metrics can be found. The result might be formalized as maxValues $\in R^n$ where n is a number of metrics collected for each VM.

$$maxValues = \begin{bmatrix} M_1 \\ M_2 \\ \vdots \\ M_n \end{bmatrix}$$

Then each metric in a corresponding metrics set is divided by the maximum number within this particular set of metrics from all VMs.

$$S = \frac{Y}{maxValues}$$

Where *Y* is a matrix with all collected metrics and *S* is a resulting matrix with all metrics being scaled to the range between 0 and 1. Now parameter ε can be chosen as 0.75 and DBSCAN algorithm can be applied to this new matrix S.

At this point any outliers in each domain can be identified. This means that some performance metrics of any identified VMs are significantly different from the others. It does not, however, mean that something is wrong with this particular VM. It might be a case when all the testing environments are idle at this time and only one or a few of them are under load due to the launching of tests. In a production model, this kind of outlying VM might be caused by increased customer activity or service. It is too early to say if this machine is anomalous or not. The next step is to check if this VM is normally under this kind of load by checking the performance metrics for the previous 4 hours. Compare the metrics that showed the VM was an outlier to the previous metrics to see if they differ by significantly. To say if a VM is anomalous or not following check is being processed: mean value and standard deviation of the metrics for the last 4 hours are calculated and the last metrics are compared in terms if they are closer than one standard deviation to the mean. If not - this VM being marked as anomalous.

The average value *μ* of the metrics can be described as follow:

$$\mu = \frac{1}{m} \sum_{i=1}^{m} X_i$$

Where $X_i$ is a set of metrics for a single VM.

The standard deviation *σ* of the metrics can be described as follow:

$$\sigma = \frac{\sqrt{\sum_{i=1}^{m} (x_i - \mu)^2}}{m}$$

Where $X_i$ is a set of metrics for a single VM and *μ* is a mean value described above.

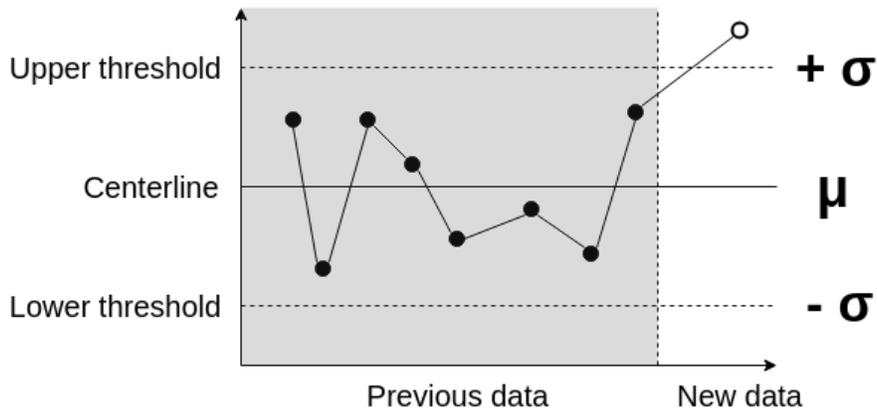
Figure 4. Intuitive graph with centerline and thresholds.

Figure 4 demonstrates an intuitive graph with a centerline, which is a mean value of all the data and a thresholds. In the presented graph one sample is outside of the thresholds so it might be marked as an anomaly.

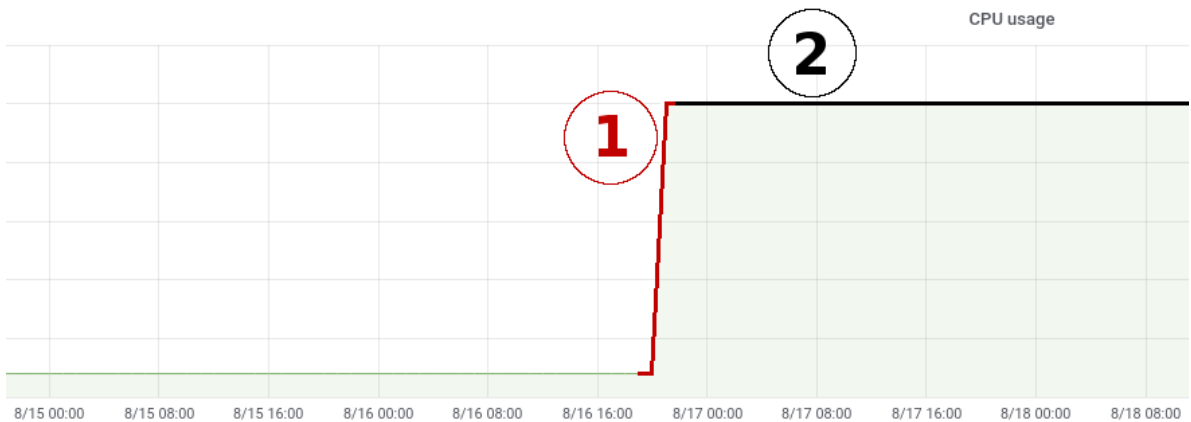
Figure 5. CPU usage grow

A good example of a case when the increase in the metric is of a normal nature is depicted on figure 5 at moment (2). At the moment (1) depicted as red on this graph a model would detect an anomaly which would need to be evaluated manually by an engineer. If an engineer would mark this case as truly anomalous then, most likely, after the fix the usage would go back to a normal level. However if the case would be marked as a normal behavior due to some legit load then at all the next iterations (2) depicted as black this host would not be marked as anomalous by the model so no human time needed to investigate the reason of the load.

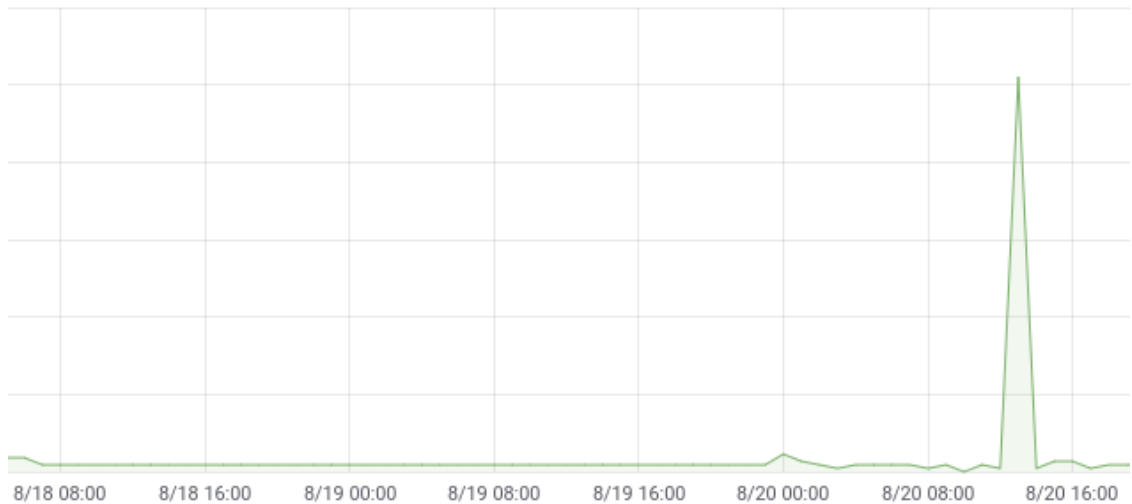

Figure 6. A spike in CPU usage

On the other hand figure 6 represents a case when a CPU usage had a huge spike. If this VM has been detected as an outlier on the previous step of the model then this kind of spike let us think of it as an unexpected behavior and could be marked as anomalous.

## 4. Experiments

This section contains results of tests that have been conducted to evaluate performance of the presented model. As a reference a standard MASF algorithm has passed the same tests and results have been compared. Tests consist of multiple anomaly injections of two categories:

- Synthetic increase of metrics for one single VM amongst other VMs running the same micro-service component at the time when all of them have been idle.

- Case opposite to the first one. When most of VMs have been under a load a micro-service on one of them have been switched off which lead to an idle state.

The following performance measures have been used:

- FP - False Positive. Means that an algorithm marked a VM as anomalous, where it was not.

- FN - False Negative. VM has been marked as normal despite its anomalous behavior.

- TP - True Positive. VM has been correctly marked as anomalous.

- TN - True Negative. VM has been correctly marked as not anomalous.

Using the measures above, the following evaluations have been used:

- Recall = TP / (TP + FN)

- Precision = TP / (FP + TP)

Altogether tests cover 177 VMs with only 4 of them being anomalous: 2 for each category of an anomaly. Table 1 and Table 2 both contain the results of the tests for the model presented in this article and MASF algorithm respectively. Left column is real labels and the top row shows results detected by each algorithm. Table 3 contains measurement results.

| Real \ Predicted | Normal | Anomalous |
|---|---|---|
| Normal | 171 | 2 |
| Anomalous | 0 | 4 |

Table 1. Tests results for the presented model

| Real \ Predicted | Normal | Anomalous |
|---|---|---|
| Normal | 155 | 18 |
| Anomalous | 2 | 2 |

Table 2. Tests results for MASF algorithm

| Algorithm | FP | FN | TP | TN | Recall | Precision |
|---|---|---|---|---|---|---|
| Presented model | 2 | 0 | 4 | 171 | 1 | 0.66 |
| MASF | 18 | 2 | 2 | 155 | 0.5 | 0.1 |

Table 3. Comparison of measures for two algorithms

Because MASF implies that data distribution will be Gaussian, whereas in most cases it is not, its predictions often determine outliers incorrectly, as could be seen in Table 2. MASF based algorithm marked 18 normal VMs as anomalous along with missing 2 anomalous VMs. The cluster based model was able to catch all 4 of anomalous VM having 100% recall. Two normal VMs have been marked as anomalous, which might be caused by a couple of scheduled functional tests running at the same time. Manual checking of each detected VM is required in order to make sure there are anomalies.

## 5. Conclusion and future work.

The presented model detects anomalous Virtual Machines within both production and testing LAB environments with good confidence. Some improvements could be done in order to have even better results in testing environments. First, this model does not consider time of day and day of week dependability of the VM load. For example: night is usually a more busy time since a lot of auto-tests are running during the night in the testing infrastructure. Some tests are being run at the same time every day. Based on this the following improvements of the model might be done:

- Analyze a detected outlier based on the same time as it was detected but for several days before. Check if this is a case when load is scheduled and planned.

- Divide the metrics used for analysis into business days vs weekend since the load might differ.

One should note that the above is not relevant to a production system because the load is always dispersed across all services and time-based spikes would affect all the VMs, so all of them should behave similarly to each other.

Another issue with the current model is that the engineer needs to manually review a detected anomalous machine in order to identify the exact reason for the anomaly or mark this case as a False Positive. Some part of this job might be automated and included into the model.

There is the possibility of using supervised learning algorithms and the presented model may help a lot in terms of collecting both positive and negative labels.

All of the improvements listed above may increase the accuracy of the model.